%% file: main.tex
\documentclass[cameraready]{Interspeech}

\usepackage{algorithm}
\usepackage{algpseudocode}
\usepackage{amsmath}
\usepackage{dsfont}
\usepackage{graphicx}
\usepackage{subcaption}

\title{MSpoofTTS: Multi-Resolution Spoof-Guided Inference for Discrete Speech Synthesis}

\author[affiliation={1}, orcid=0009-0008-2616-6590]{Junchuan}{Zhao*}
\author[affiliation={2}, orcid=0009-0001-5755-8692]{Minh Duc}{Vu*}
\author[affiliation={1}, orcid=0000-0002-0123-1260]{Ye}{Wang}

\address{
    $^1$ School of Computing, National University of Singapore, Singapore \\
    $^2$ Department of Statistics \& Data Science, National University of Singapore, Singapore
}

\email{junchuan@u.nus.edu, minhduc.vu@u.nus.edu, dcswangy@nus.edu.sg}

\keywords{speech synthesis, spoof detection, neural codec language models}

\usepackage{comment}

\begin{document}

\maketitle

\begingroup
\renewcommand\thefootnote{*}
\footnotetext{These authors contributed equally to this work.}
\endgroup

\begin{abstract}
    \input{sections/abstract}
    
\end{abstract}
\input{sections/intro}
\input{sections/method}
\input{sections/exp_setup}
\input{sections/exp_result}
\input{sections/conclusion}



\bibliographystyle{IEEEtran}
\bibliography{mybib}

\end{document}

%% file: sections/abstract.tex
Neural codec language models enable high-quality discrete speech synthesis, yet their inference remains vulnerable to token-level artifacts and distributional drift that degrade perceptual realism. Rather than relying on preference optimization or retraining, we propose MSpoof-TTS, a training-free inference framework that improves zero-shot synthesis through multi-resolution spoof guidance. We introduce a Multi-Resolution Token-based Spoof Detection framework that evaluates codec sequences at different temporal granularities to detect locally inconsistent or unnatural patterns. We then integrate the spoof detectors into a hierarchical decoding strategy, progressively pruning low-quality candidates and re-ranking hypotheses. This discriminator-guided generation enhances robustness without modifying model parameters. Experiments validate the effectiveness of our framework for robust and high-quality codec-based speech generation. Audio samples\footnote{\url{https://danny-nus.github.io/MSpoofTTS.github.io/}} and code\footnote{\url{https://github.com/Danny-NUS/MSpoofTTS}} are available.

%% file: sections/intro.tex
\section{Introduction}
Neural codec language models have recently become a practical and effective approach for zero-shot speech synthesis \cite{tan2024naturalspeech, ju2024naturalspeech, du2024cosyvoice, ye2025llasa, meng2025autoregressive, wangmaskgct, song2025ella, chen2025f5, zhao2025prosody, liang2026segment}. As these systems continue to improve, robustness during decoding has become an important concern alongside expressiveness and controllability \cite{xie2025towards, choi2025personalized, zhao2024sintechsvs, alwaisi2024advancements, zhao2025spsinger}, particularly in long-form and zero-shot generation settings. By modeling speech as sequences of discrete codec tokens with autoregressive or transformer architectures, these systems streamline the synthesis pipeline and naturally adopt scalable decoding strategies from large language models. In practice, however, generation in the discrete token space can be fragile. Small inconsistencies at the token level may accumulate during autoregressive decoding, resulting in audible artifacts, locally unnatural transitions, or a gradual drift away from natural speech characteristics \cite{zhang2024speechalign}. Since next-token prediction objectives do not explicitly constrain such behavior, these issues are often difficult to detect or correct during inference.

To mitigate these decoding instabilities, existing methods generally fall into two categories. One line of work introduces additional supervision or reward-driven objectives to better align model outputs with perceptual or distributional goals \cite{zhang2024speechalign, hu2024robust, gao2025differentiable, chen2024enhancing, zhao2026comelsinger}. For example, SpeechAlign \cite{zhang2024speechalign} reduces the mismatch between golden and synthesized codec tokens through preference-based optimization. Building on similar ideas, \cite{hu2024robust} leverages human feedback signals as training supervision to improve perceptual quality, while \cite{gao2025differentiable} integrates differentiable reward signals directly into the optimization objective to refine generation behavior. In addition, \cite{chen2024enhancing} introduces an auxiliary reverse inference objective to enhance robustness under distributional shift. 
Although effective, these approaches often require retraining, iterative optimization, or carefully curated data, which increases computational cost and system complexity. 

Another line of work focuses on decoding-time adjustments, including repetition control, alignment constraints, or modified sampling strategies \cite{chen2024vall, song2025ella, han2024vall}. VALL-E 2 \cite{chen2024vall} introduces repetition-aware sampling to mitigate degeneration during autoregressive decoding. ELLA-V \cite{song2025ella} improves stability by refining decoding strategies under long-context and zero-shot settings. VALL-E R \cite{han2024vall} enhances robustness through inference-time alignment constraints and decoding refinements. These methods are simple to apply and avoid retraining, but they primarily target specific failure patterns rather than explicitly assessing whether the generated token sequence remains globally consistent or locally natural.

A related idea has been explored in controllable text generation, where external classifiers are incorporated directly during decoding to guide model outputs without retraining the base language model \cite{liu2021dexperts, dathathri2019plug}. Instead of optimizing the generator through reinforcement learning, these approaches influence generation at inference time by adjusting token probabilities or re-ranking candidate sequences based on auxiliary evaluation signals. Such strategies demonstrate that generation behavior can be steered through an explicit evaluation mechanism applied during decoding, rather than by modifying model parameters. Motivated by this perspective, we introduce a spoof detection model as an authenticity evaluator and integrate it into the decoding process to guide discrete speech generation.

To this end, spoofed and deepfake speech detection provides a natural foundation. Detection of synthetic or manipulated speech has been extensively studied in the continuous audio domain \cite{sanchez2015toward, zhang2021one, wang2024generalizable, sun2025contrastive, javed2022voice}, with large-scale benchmarks such as the ASVspoof challenge series \cite{wu2017asvspoof, todisco2019asvspoof, yamagishi2021asvspoof, wang2024asvspoof} driving substantial progress in waveform-level countermeasure design.
More recently, codec-based deepfake datasets have revealed additional challenges introduced by neural codec synthesis \cite{wu2024codecfake, chen2025codecfake+, du2025codecfake}. However, existing spoof detection systems are primarily formulated on reconstructed audio signals and are designed for post-hoc classification. They do not operate on discrete codec token sequences, nor are they intended to guide generation during decoding.

\begin{figure*}[t]
    \centering
    \includegraphics[width=0.9\linewidth]{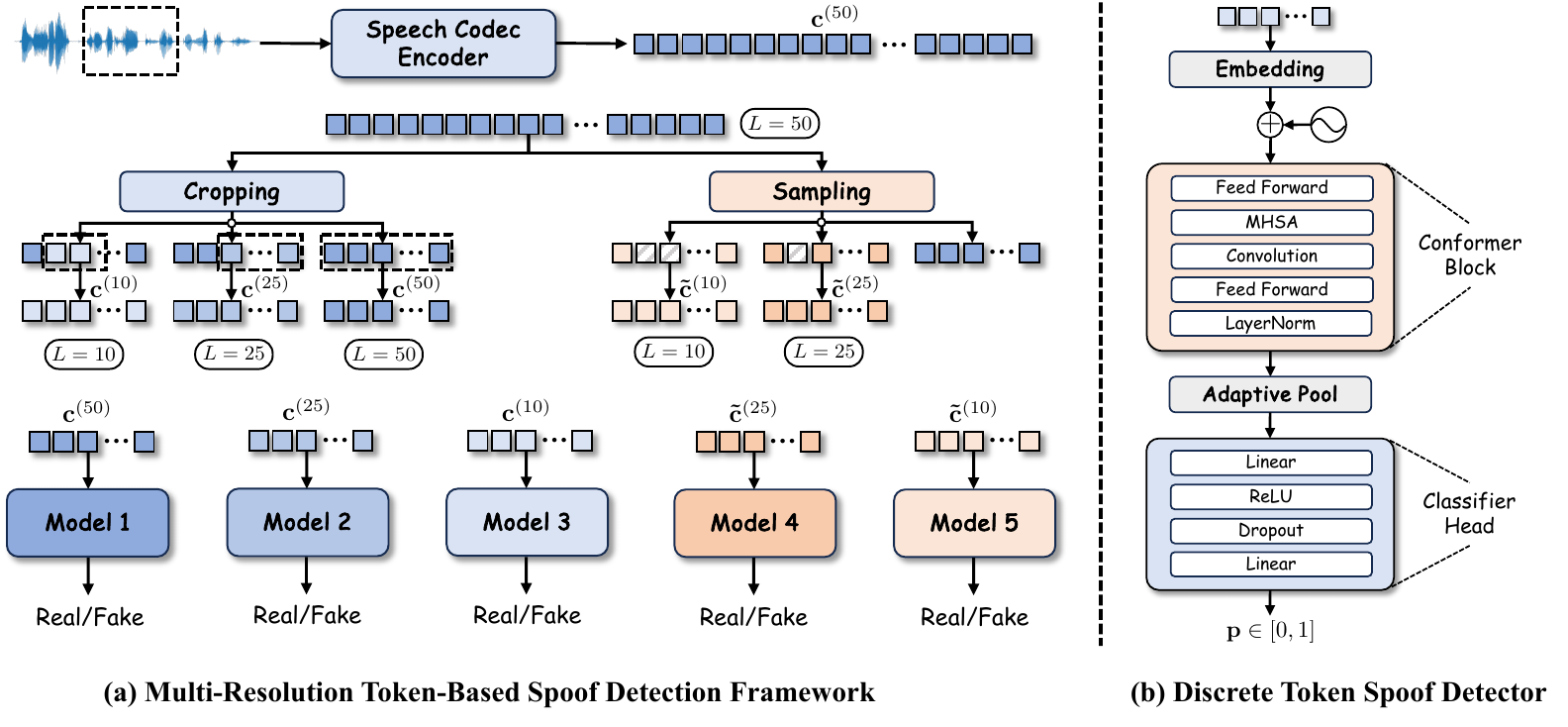}
    \caption{Overview of multi-resolution token-based spoof detection framework. (a) Construction of token sequences at multiple temporal resolutions for training separate real/fake detectors. (b) Conformer-based discrete token spoof detector architecture.}
    \label{fig:overall}
\end{figure*}

Building on this perspective, we propose MSpoof-TTS, a framework that integrates a separately trained multi-resolution token-level spoof detector into the decoding process for codec-based speech synthesis, while keeping the underlying language model fixed. 
The main contributions of this work are as follows:
\begin{itemize}
\item We extend spoof detection to the token level by introducing a multi-resolution authenticity modeling approach tailored to discrete codec sequences.
\item We develop an inference-time decoding strategy that leverages spoof-based authenticity scores for candidate pruning and reranking, without retraining the base codec language model.
\item We demonstrate consistent improvements in perceptual quality and robustness across diverse decoding configurations.
\end{itemize}

%% file: sections/method.tex
\section{Method}
\subsection{Overview}

\begin{figure}[t]
    \centering

    \begin{subfigure}{0.485\linewidth}
        \centering
        \includegraphics[width=\linewidth]{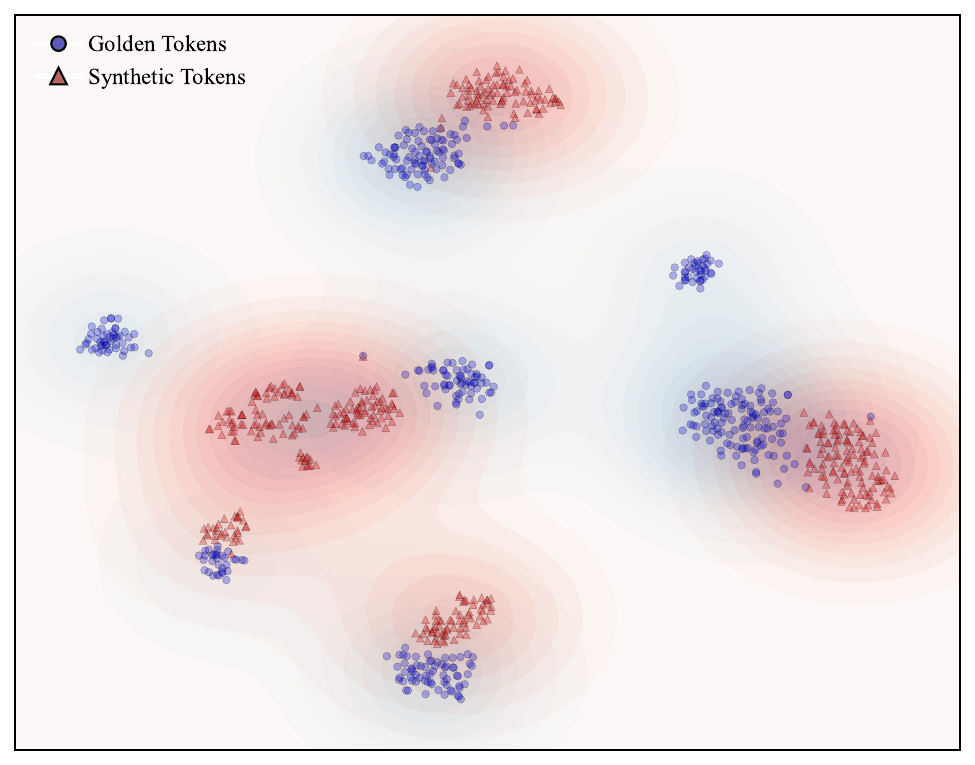}
        \caption{Full Utterance}
    \end{subfigure}
    \hspace{0.01\linewidth}
    \begin{subfigure}{0.485\linewidth}
        \centering
        \includegraphics[width=\linewidth]{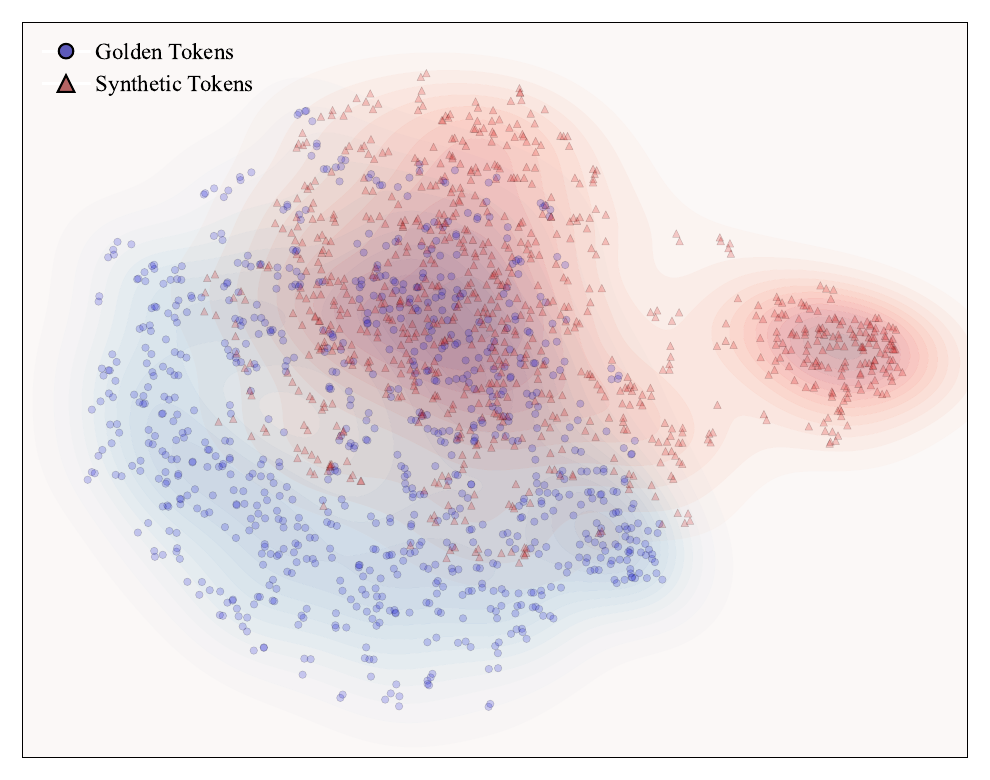}
        \caption{Segment Length = 50}
    \end{subfigure}

    \vspace{2pt}

    \begin{subfigure}{0.485\linewidth}
        \centering
        \includegraphics[width=\linewidth]{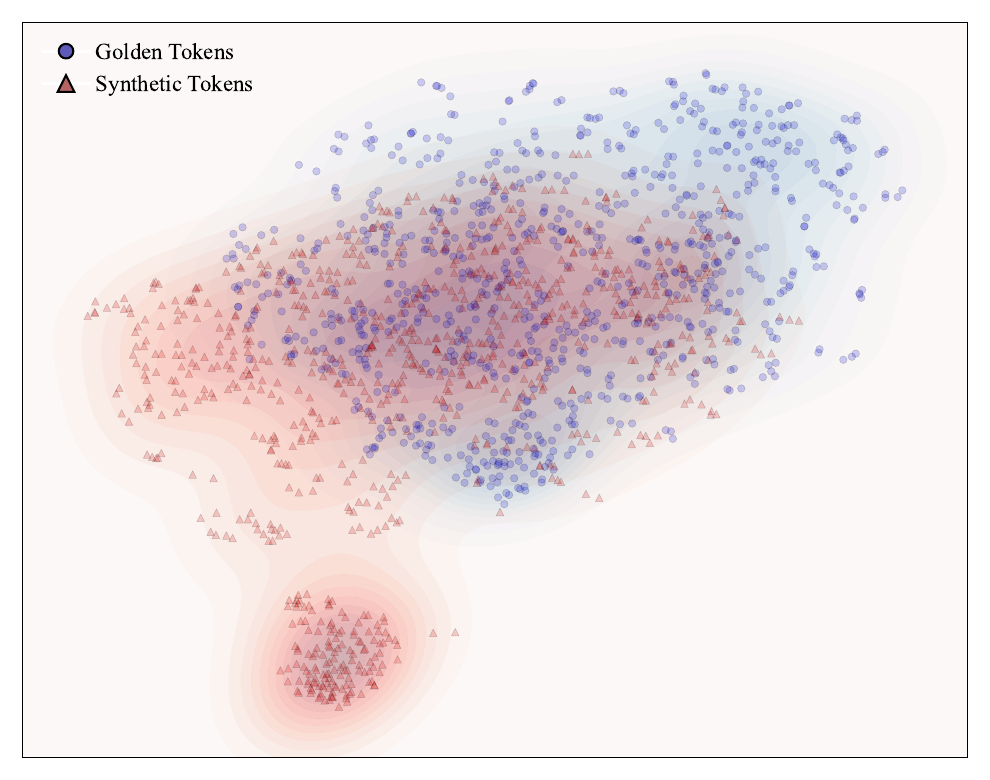}
        \caption{Segment Length = 25}
    \end{subfigure}
    \hspace{0.01\linewidth}
    \begin{subfigure}{0.485\linewidth}
        \centering
        \includegraphics[width=\linewidth]{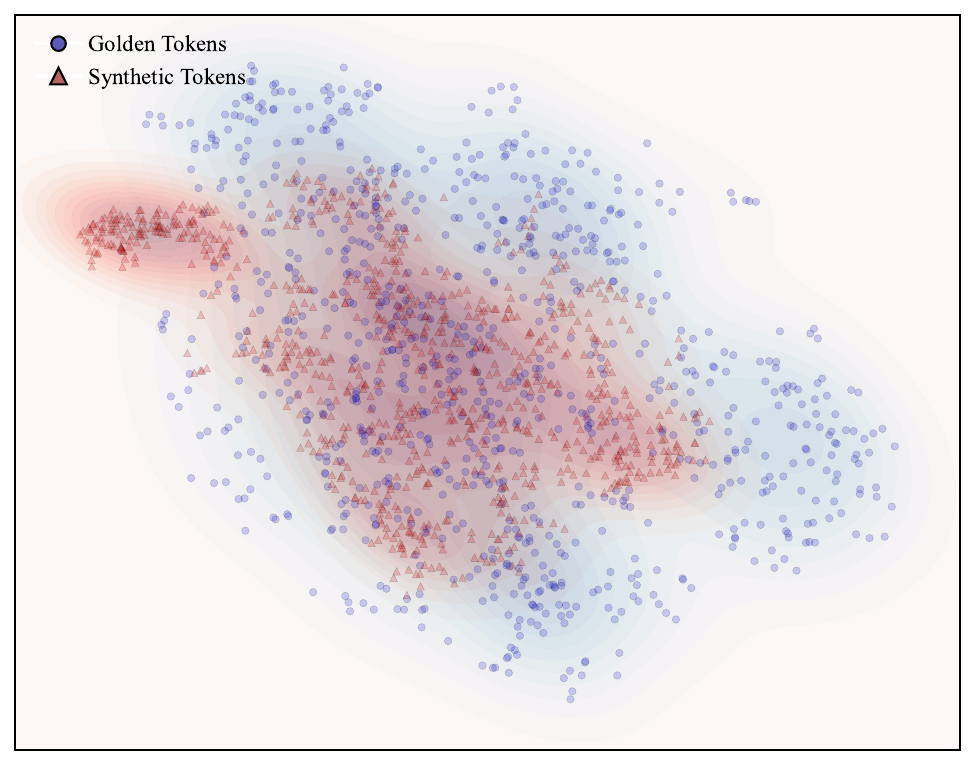}
        \caption{Segment Length = 10}
    \end{subfigure}

    \caption{t-SNE visualization of embedding distributions under different segment lengths.}
    \label{fig:tsne_all}
\end{figure}

Pretrained codec-based language models have demonstrated strong capability in zero-shot speech synthesis by modeling discrete codec token sequences autoregressively. In this work, we adopt NeuTTS\footnote{\url{https://github.com/neuphonic/neutts}} \cite{julian2025finite}, a pretrained codec-based TTS system, as the base generator, whose parameters remain fixed throughout our framework. While such models are trained to predict next tokens conditioned on ground-truth codec sequences, inference relies on previously generated tokens. This shift introduces a discrepancy between training and decoding, under which synthesized token sequences may progressively deviate from the natural codec distribution, as previously observed in \cite{zhang2024speechalign}.

Following this approach, token representations are obtained by mean pooling over the temporal dimension to aggregate token embeddings across multiple temporal resolutions. As illustrated in Figure~\ref{fig:tsne_all}, we observe a consistent trend, where the statistical gap between golden and synthesized codec tokens appears not only at the utterance level but also across multiple segment lengths, including 50-token, 25-token, and 10-token segments. This indicates that the distribution mismatch manifests at different temporal resolutions, suggesting that local token-level deviations accumulate during decoding and contribute to perceptual degradation.

Rather than modifying the pretrained TTS model to compensate for this mismatch, we explicitly model the discrepancy through a spoof detection framework. Specifically, we train discriminators (spoof detection models) to distinguish the golden (ground-truth) codec sequences from synthetic ones, using token segments constructed at multiple temporal scales. The resulting authenticity scores are then incorporated into an inference-time guidance strategy that evaluates candidate sequences during decoding. By progressively pruning low-quality candidates and re-ranking hypotheses based on spoof scores, the proposed framework steers generation toward more natural and robust outputs without altering the underlying codec language model.

\subsection{Multi-Resolution Token-Level Spoof Detection}\label{subsec:MRSD}
To explicitly model the distribution gap between golden and synthesized codec tokens, we propose a multi-resolution token-level spoof detection framework as illustrated in Figure~\ref{fig:overall}. Given an utterance-level codec sequence $\mathbf{c} = \{c_1, \dots, c_T\}$, we construct token segments using two complementary resolution strategies to expose discrepancies across temporal and structural scales.

We first vary the temporal span by extracting contiguous subsequences with lengths $L \in \{10, 25, 50\}$. Short segments emphasize fine-grained and local transition dynamics, whereas longer segments capture broader contextual coherence and higher-level structural dependencies. Modeling multiple segment lengths enables the detector to approximate distributional characteristics across diverse temporal horizons.

In addition, inspired by multi-scale discriminators in neural vocoders \cite{kong2020hifi, leebigvgan}, we introduce a resolution-based skip sampling strategy with downsampling rates $r \in \{1, 2, 5\}$. When $r=1$, the sequence remains at its native resolution, corresponding to the full 50-token span. Larger $r$ values subsample the sequence to form coarser token representations, encouraging the detector to uncover structural inconsistencies that may not be evident at the original granularity. While temporal cropping enhances coverage across time spans, skip sampling perturbs token resolution to probe structural regularity from a complementary perspective.

Each token segment is processed by an embedding layer followed by stacked Conformer \cite{gulati2020conformer} blocks to jointly capture local correlations and long-range dependencies. The resulting sequence representation is aggregated via adaptive pooling and passed to a lightweight classifier to predict the probability of the segment being real or synthetic. The detector is trained with the binary cross-entropy (BCE) objective and optimized independently of the TTS backbone. The architecture is shared across all resolution settings, while parameters are learned separately for each configuration. Specifically, we train five models: $\mathcal{M}_{50}$, $\mathcal{M}_{25}$, and $\mathcal{M}_{10}$ for contiguous cropping with segment lengths $L=50,25,10$, respectively, and $\mathcal{M}_{50 \leftarrow 25}$ and $\mathcal{M}_{50 \leftarrow 10}$ for skip-sampled variants obtained by downsampling 50-token segments with rates $r=2$ and $r=5$, respectively.

\begin{algorithm}[t]
\caption{Entropy-Aware Sampling (EAS)}
\label{alg:eas}
\begin{algorithmic}[1]
\Require Text $x$, pre-trained AR model $\theta_{AR}$, prefix $c_{<t}$;
top-$p$ threshold $v$, cluster size $k_e$, memory window size $W$; 
$\alpha,\beta,\gamma$
\Ensure Generated tokens $c_{1:T}$
\State $\mathcal{M}\gets\emptyset$
\For{$t=1,2,\dots,T$}
    \State $\mathbf{s}_t \gets p(\cdot \mid x,c_{<t}; \theta_{AR})$
    \ForAll{$(i,r,a)\in\mathcal{M}$}
        \State $\pi_t(j)\gets\min(\gamma,\;\sum_{(i,r,a)\in\mathcal{M},\ i=j}\alpha\cdot\frac{1}{1+r}\cdot\beta^{a})$
    \EndFor
    \State $\mathbf{s}'_t \gets \mathbf{s}_t - \boldsymbol{\pi}_t$
    \State $c_t \gets \mathrm{NucleusSample}(\mathbf{s}'_t;\,v)$
    \State $\mathcal{M}\gets \{(i,r,a+1):(i,r,a)\in\mathcal{M},\ a+1\le W\}$
    \State $\mathcal{K}_t \gets \mathrm{TopK}(\mathbf{s}'_t,k_e) \cup \{c_t\}$
    \For{$r=1, \cdots, |\mathcal{K}_t|$}
        \State $\mathcal{M}\gets \mathcal{M}\cup\{(\mathcal{K}_t[r],r,0)\}$
    \EndFor
    \State $c_{<t+1}\gets c_{<t}  \cup \{c_t\}$
\EndFor
\end{algorithmic}
\end{algorithm}

\begin{algorithm}[t]
\caption{Hierarchical Sampling with Progressive Discriminator Pruning}
\label{alg:hier-sampling}
\begin{algorithmic}[1]
\Require $x$, $\theta_{AR}$;
discriminators $\mathcal{M}_{10},\mathcal{M}_{25},\mathcal{M}_{50}$;
$\mathcal{M}_{50\leftarrow25},\mathcal{M}_{50\leftarrow10}$;
warmup lengths $L_w$;
stage lengths $(L_1,L_2,L_3)$;
beams $(B_0,B_1,B_2)$;
max length $L_{\max}$;
rank weights $(w_{50},w_{25},w_{10})$
\Ensure $y$

\State $y \gets x$

\Statex // Warmup using EAS
\State $y \gets y \cup \mathrm{EAS}(y,L_w;\theta_{AR})$

\While{$|y| < L_{\max}$}

\Statex // Stage 1: generate $B_0$ candidates up to $L_1$
\State $\mathbf{B} \gets \{\, \mathrm{EAS}(y,L_1;\theta_{AR}) \,\}_{b=1}^{B_0}$

\Statex // Prune to top $B_1$ by $\mathcal{M}_{10}$
\ForAll{$b_i \in \mathbf{B}$}
    \State $s_i \gets \mathcal{M}_{10}(b_i)$
\EndFor
\State Keep top $B_1$ beams ranked by $s_i$

\Statex // Stage 2: extend remaining beams to $L_2$
\ForAll{$b_i \in \mathbf{B}$}
    \State $b_i \gets b_i \cup \mathrm{EAS}(b_i,L_2 - L_1;\theta_{AR}) $
\EndFor

\Statex // Prune to top $B_2$ by $\mathcal{M}_{25}$
\ForAll{$b_i \in \mathbf{B}$}
    \State $s_i \gets \mathcal{M}_{25}(b_i)$
\EndFor
\State Keep top $B_2$ beams ranked by $s_i$

\Statex // Stage 3: extend to $L_3$
\ForAll{$b_i \in \mathbf{B}$}
    \State $b_i \gets b_i \cup \mathrm{EAS}(b_i,L_3 - L_2;\theta_{AR}) $
\EndFor

\Statex // Rank aggregation for final selection
\ForAll{$b_i \in \mathbf{B}$}
    \State $s_{50}[i] \gets \mathcal{M}_{50}(b_i)$
    \State $s_{25}[i] \gets \mathcal{M}_{50\leftarrow25}(b_i)$
    \State $s_{10}[i] \gets \mathcal{M}_{50\leftarrow10}(b_i)$
\EndFor

\State $\mathbf{r}_{50},\mathbf{r}_{25},\mathbf{r}_{10} \gets \mathrm{Rank}(\mathbf{s}_{50},\mathbf{s}_{25},\mathbf{s}_{10})$
\State $R[i] \gets w_{50}r_{50}[i] + w_{25}r_{25}[i] + w_{10}r_{10}[i]$
\State $b^\star \gets \arg\min_i R[i]$

\State $y \gets y \cup b^\star$

\EndWhile

\Return $y$
\end{algorithmic}
\end{algorithm}

\subsection{Hierarchical Spoof-Guided Sampling}
To improve the decoding robustness of autoregressive codec language models, we propose a hierarchical spoof-guided sampling framework that integrates Entropy-Aware Sampling (EAS) with the multi-resolution token-level spoof detection introduced in Section~\ref{subsec:MRSD}. We first adopt Entropy-Aware Sampling (EAS), shown in Algorithm~\ref{alg:eas}, as the base decoding strategy. EAS is adapted from repetition-aware sampling (RAS) proposed in VALL-E 2~\cite{chen2024vall}, which penalizes recently generated tokens to mitigate looping behavior. However, conventional RAS relies on heuristic repetition counting and may over-suppress high-probability tokens, resulting in unstable decoding or excessive randomness. In contrast, EAS maintains a memory buffer that records competitive candidate tokens along with their rank positions and temporal age. The token-level penalty is modulated by inverse rank weighting and exponential temporal decay, with a clipping mechanism to prevent over-penalization. The adjusted distribution is then used for nucleus sampling, introducing entropy regularization while preserving distributional diversity.

Building upon EAS, we introduce a hierarchical spoof-guided pruning strategy, summarized in Algorithm~\ref{alg:hier-sampling}. After generating an initial warmup segment of length $L_w$ to stabilize early decoding, the algorithm proceeds iteratively. At each iteration, $B_0$ candidate continuations are generated from the current prefix using EAS up to length $L_1$. These candidates are evaluated by the short-span discriminator $\mathcal{M}_{10}$, and only the top $B_1$ beams are retained. The remaining beams are then extended to length $L_2$ using EAS and pruned by the mid-range discriminator $\mathcal{M}_{25}$ to keep the top $B_2$ candidates. Finally, the surviving beams are extended to length $L_3$, forming complete candidate segments.

For final selection, we perform multi-resolution rank aggregation rather than relying on a single discriminator score. Each candidate segment is evaluated by the long-span discriminator $\mathcal{M}_{50}$ and its sampled variants $\mathcal{M}_{50\leftarrow25}$ and $\mathcal{M}_{50\leftarrow10}$. The ranking positions across resolutions are combined using weighted aggregation, and the candidate with the best aggregated score is appended to the decoded sequence. This procedure repeats until reaching the maximum decoding length. The proposed coarse-to-fine, multi-resolution pruning improves decoding stability and structural consistency without modifying or retraining the AR backbone.

%% file: sections/exp_setup.tex
\section{Experiments}

\subsection{Datasets}

For spoof detection training, we use the LibriTTS \cite{zen2019libritts} training split (approximately 100 hours of clean read English speech). LibriTTS is a multi-speaker corpus with careful segmentation and noise filtering. For each ground-truth utterance, we generate three synthetic counterparts using the same transcript but a different reference utterance from the same speaker under the default inference scheme. All real and synthetic waveforms are converted into discrete token sequences using the default speech tokenizer (NeuCodec).

For evaluation, experiments are conducted on both LibriSpeech \cite{panayotov2015librispeech} and LibriTTS (test split), which provide diverse speakers and clean TTS-style speech, respectively. These datasets enable evaluation under standard speech conditions using WER, speaker similarity, and perceptual quality metrics. To further assess robustness under challenging phonetic patterns, we additionally evaluate on the TwistList benchmark \cite{loakman-etal-2023-twistlist}, which consists of linguistically constructed tongue twisters characterized by dense alliteration and repetitive phoneme patterns. We use the official test split and synthesize speech using the same conditioning and decoding pipeline as in the standard evaluation.

\subsection{Implementation Details}
The spoof detection models use $d_{\text{model}}=256$, 8 attention heads, 4 Transformer layers, and a feedforward dimension of 1024, with dropout set to 0.1. The discriminator is trained with AdamW using a learning rate of $1\times10^{-4}$ and weight decay of $1\times10^{-4}$ on a single NVIDIA L40S GPU. For decoding, the default baseline adopts top-$k$ sampling with $k=50$ and temperature 1.0. Repetition-Aware Sampling (RAS) uses top-$k=50$, top-$p=0.8$, window size $W=25$, and repetition penalty $\tau_r=0.1$. Entropy-Aware Sampling (EAS) follows Algorithm~\ref{alg:eas} with cluster size $k_e=3$, memory window $W=15$, and hyperparameters $\alpha=0.2$, $\beta=0.7$, and $\gamma=0.8$, while using top-$p=0.8$, top-$k=50$, and temperature 1.0. 
Hierarchical decoding follows Algorithm~\ref{alg:hier-sampling} with warm-up length $L_w=20$, stage lengths $(L_1, L_2, L_3)=(10, 25, 50)$, and beam sizes $(B_0, B_1, B_2)=(8, 5, 3)$. Ranking weights $(w_{50}, w_{25}, w_{10})$ are set equally. Hierarchical RAS and EAS inherit the corresponding sampling hyperparameters.

\subsection{Evaluation Metrics}

For objective evaluation, we report ASR-based intelligibility, speaker similarity, perceptual quality, and spoof detection performance. Intelligibility is measured using Word Error Rate (WER), computed with Whisper-large-v3 \cite{radford2023robust}\footnote{\url{https://huggingface.co/openai/whisper-large-v3}} for all decoding schemes. Speaker similarity is evaluated using the WavLM \cite{chen2022wavlm}\footnote{\url{https://huggingface.co/microsoft/wavlm-base-plus-sv}} model by computing cosine similarity between L2-normalized embeddings of the generated utterance and another real utterance from the same speaker. Perceptual quality is assessed using neural MOS estimators, including NISQA \cite{mittag2021nisqa} and MOSNet \cite{lo2019mosnet}, both reflecting overall speech quality.

For spoof detection evaluation, we report Accuracy, Macro-F1 (threshold = 0.5), and AUROC. As the spoof detiection models is primarily used for ranking candidate token sequences, AUROC is considered the main indicator of discriminative capability. In addition, subjective evaluation is conducted using mean opinion scores for naturalness (MOS-N), quality (MOS-Q), and speaker similarity (SMOS). 15 participants rate audio samples from different inference methods on a 1–5 scale, where higher scores indicate better perceptual quality.

%% file: sections/exp_result.tex
\section{Results}

\begin{table*}[t]
\centering
\caption{Objective evaluation on LibriTTS and LibriSpeech. HierEAS (MSpoofTTS) uses Hierarchical Spoof-Guided Sampling with EAS, while HierRAS replaces EAS with RAS under the same framework. Original refers to the vanilla top-k sampling method. The best results are highlighted in \textbf{bold}, and the second-best results are \underline{underlined}.}
\begin{tabular}{lcccccccc}
\toprule
& \multicolumn{4}{c}{LibriSpeech} 
& \multicolumn{4}{c}{LibriTTS} \\
\cmidrule(lr){2-5} \cmidrule(lr){6-9}
Inference Scheme 
& WER $\downarrow$ & SIM $\uparrow$ & NISQA $\uparrow$ & MOSNET $\uparrow$
& WER $\downarrow$ & SIM $\uparrow$ & NISQA $\uparrow$ & MOSNET $\uparrow$ \\
\midrule
Ground Truth 
& 0.0337 & 0.915 & 4.620 & 4.5259 
& 0.0430 & 0.904 & 4.593 & 4.4766\\
\midrule
Original 
& 0.0694 & 0.894 & 4.462 & 4.3418 
& 0.0715 & 0.872 & 4.397 & 4.1879 \\

RAS 
& 0.0641 & \textbf{0.905} & 4.553 & 4.2772
& 0.0657 & 0.880 & 4.425 & 4.2565\\

EAS 
& \underline{0.0576} & \underline{0.902} & 4.571 & 4.3298
& 0.0672 & 0.883 & 4.408 & 4.1937 \\

HierRAS
& 0.0591 & 0.902 & \underline{4.596} & \underline{4.3486}
& \textbf{0.0628} & \textbf{0.886} & \underline{4.520} & \underline{4.3277}\\

HierEAS
& \textbf{0.0532} & 0.901 & \textbf{4.602} & \textbf{4.4158}
& \underline{0.0633} & \underline{0.883} & \textbf{4.562} & \textbf{4.3409}\\

\bottomrule
\end{tabular}
\label{tab:standard_results}
\end{table*}

\subsection{Evaluation on Standard Benchmarks}
We compare several inference strategies, including the default top-k sampling decoder (Original), repetition-aware sampling (RAS), entropy-aware sampling (EAS), and their hierarchical extensions (HierRAS and HierEAS). HierRAS and HierEAS integrate the proposed hierarchical spoof-guided sampling framework with RAS and EAS, respectively, with HierEAS corresponding to the proposed MSpoofTTS method.

As shown in Table~\ref{tab:standard_results}, both EAS and RAS consistently outperform the original top-k sampling baseline across LibriSpeech and LibriTTS, confirming that entropy- and repetition-aware decoding mitigates degeneration and improves token-level selection. Incorporating hierarchical spoof-guided ranking yields further improvements in most settings. Notably, hierarchical variants demonstrate more consistent gains in perceptual quality metrics, including NISQA and MOSNET, across both datasets. These results suggest that multi-resolution discriminator guidance enhances structural consistency beyond local decoding refinements.

In contrast, improvements in WER and SIM are comparatively modest. Given the already strong lexical accuracy and speaker consistency of the baseline decoder, substantial gains in these metrics are naturally limited. Importantly, hierarchical sampling preserves competitive WER and SIM performance while improving perceptual quality, indicating that discriminator-guided ranking enhances generation quality without compromising intelligibility or speaker identity. Overall, HierEAS (MSpoofTTS) achieves the best or second-best performance across most metrics and datasets, supporting the effectiveness of the proposed method.

\subsection{Evaluation under Challenging Conditions}
\begin{table}[t]
\centering
\caption{Objective evaluation on TwistList test set.}
\begin{tabular}{lccccc}
\toprule
Methods & WER $\downarrow$ & SIM $\uparrow$ & NISQA $\uparrow$ & MOSNET $\uparrow$\\
\midrule
Original & 0.1654 & 0.871 & 4.459 & 3.8692\\
RAS & 0.1572 & 0.820 & 4.477 & 3.8754\\
EAS & \textbf{0.1433} & \underline{0.876} & 4.465 & 3.9265 \\
HierRAS & 0.1702 & 0.802 & \underline{4.496} & \underline{3.9566}\\
HierEAS & \underline{0.1531} & \textbf{0.877} & \textbf{4.513} & \textbf{3.9802}\\
\bottomrule
\end{tabular}
\label{tab:twist_results}
\end{table}

Table~\ref{tab:twist_results} reports results on the TwistList test set, which consists of phonetically dense tongue-twister utterances and serves as a stress test for decoding robustness. As expected, WER values are higher across all methods compared to standard benchmarks, reflecting the increased difficulty of this dataset.

The overall trend aligns with observations on LibriTTS and LibriSpeech: entropy- and repetition-aware decoding (EAS and RAS) consistently outperform the original top-k sampling baseline in most metrics. In particular, EAS achieves the lowest WER on TwistList, slightly surpassing RAS, suggesting that entropy-aware regularization better balances repetition control with legitimate phonetic recurrence under challenging conditions. While hierarchical variants do not always obtain the lowest WER in this setting, HierEAS maintains competitive intelligibility and achieves the best perceptual quality scores in both NISQA and MOSNET. Notably, the relative increase in WER remains moderate, indicating that the proposed method preserves lexical accuracy even under highly constrained phonetic structures.

\subsection{Subjective Listening Tests}
\begin{figure}[t]
    \centering
    \includegraphics[width=0.9\linewidth]{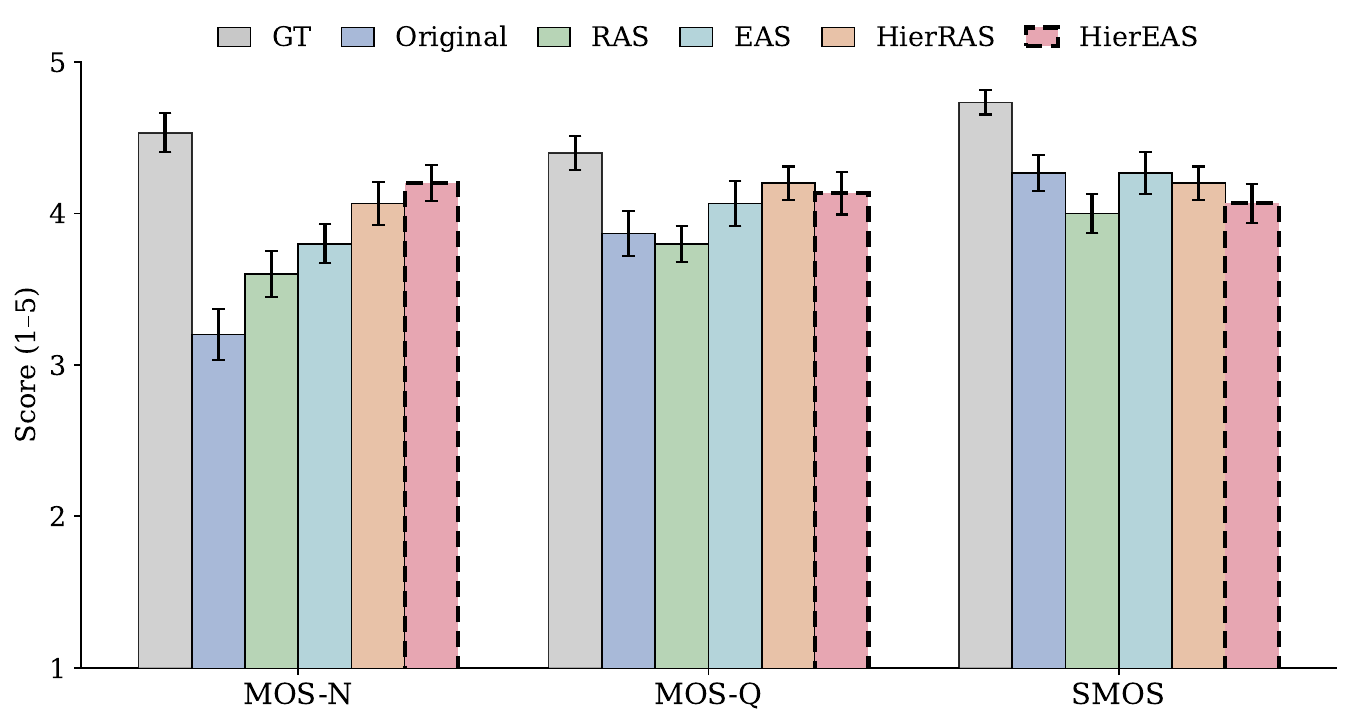}
    \caption{Subjective evaluation of different inference strategies measured by MOS-N (naturalness), MOS-Q (quality), and SMOS (similarity).}
    \label{fig:mos}
\end{figure}

We conduct a subjective evaluation across different inference strategies, reporting MOS-N, MOS-Q, and SMOS to measure naturalness, overall speech quality, and speaker similarity, respectively. As shown in Figure~\ref{fig:mos}, hierarchical spoof-guided decoding consistently achieves higher perceptual scores than the baseline methods. In particular, the hierarchical variants (HierRAS and HierEAS) demonstrate clear improvements in naturalness (MOS-N) over their non-hierarchical counterparts, suggesting that spoof-aware hierarchical decoding helps mitigate unnatural token patterns during generation. Similar trends are observed for overall quality (MOS-Q), where the hierarchical strategies achieve slightly higher scores, although the improvement remains modest. For speaker similarity (SMOS), all methods maintain consistently high scores. Although our approach does not obtain the highest SMOS, the results indicate that the proposed inference strategy preserves speaker identity while improving perceptual naturalness and quality.

\subsection{Analysis of the Multi-Resolution Spoof Detectors}
\begin{table}[t]
\centering
\caption{Evaluation of the multi-resolution spoof detection models on the LibriTTS test-clean split.}
\begin{tabular}{lcccc}
\toprule
Model & Resolution & AUC & Acc & Macro-F1 \\
\midrule
Model 1        & 50       & 0.9199 & 0.8616 & 0.8180 \\
Model 2        & 25       & 0.8298 & 0.8034 & 0.6892 \\
Model 3        & 10       & 0.6842 & 0.7506 & 0.5336 \\
Model 4        & 50$\rightarrow$25 & 0.8346 & 0.7935 & 0.5973 \\
Model 5        & 50$\rightarrow$10 & 0.6914 & 0.7456 & 0.5952 \\
\bottomrule
\end{tabular}
\label{tab:disc_results}
\end{table}

Table~\ref{tab:disc_results} reports the performance of the five multi-resolution spoof detection models on LibriTTS test-clean. A clear dependence on temporal context length is observed: the full-resolution discriminator with $L=50$ achieves the strongest performance, indicating that longer token sequences provide richer structural cues for distinguishing real from synthetic speech.

Although performance declines for shorter segments ($L=25$ and $L=10$), they retain meaningful discriminative capability, suggesting that local token-level irregularities remain detectable. The strided variants derived from $L=50$ maintain moderate ranking ability, indicating that structural signals persist under resolution perturbation. These results highlight complementary roles across temporal scales and provide empirical support for the hierarchical multi-resolution ranking strategy adopted in our inference framework.

%% file: sections/conclusion.tex
\section{Conclusion}
We present MSpoofTTS, a training-free framework that improves discrete speech synthesis through hierarchical spoof-guided inference. We use multi-resolution spoof detectors to guide decoding and suppress locally inconsistent codec token patterns, without modifying the pretrained speech language model. Experiments on LibriTTS and LibriSpeech show consistent improvements in perceptual quality while maintaining competitive intelligibility and speaker similarity. Evaluation on the challenging TwistList dataset further demonstrates robustness under highly repetitive phonetic structures. Subjective listening tests show improved perceptual naturalness without degrading speaker identity, supporting spoof-guided inference as an effective approach to enhancing neural codec-based speech generation.